\documentclass[reprint,superscriptaddress,amsmath,amssymb,aps,prb]{revtex4-2}

\usepackage{graphicx}% Include figure files
\usepackage{dcolumn}% Align table columns on decimal point
\usepackage{bm}% bold math
\usepackage{color, colortbl}% Color fonts
\usepackage{soul} % strikethrough text

\usepackage{siunitx}
\usepackage{textcomp, gensymb}
\begin{document}

\title{Geometric Control of Visible Emitter Creation in Hexagonal Boron Nitride by Oblique Ion Irradiation}

\author{Sagar Chowdhury}
\affiliation{Department of Physics \& Astronomy, University of Iowa, Iowa City, IA 52242, United States.}

\author{Bhaveshkumar Kamaliya}
\affiliation{Department of Materials Science and Engineering, McMaster University, Hamilton, Ontario L8S 4L8, Canada.}

\author{Ramachandra Bangari}
\affiliation{Department of Physics \& Astronomy, University of Iowa, Iowa City, IA 52242, United States.}

\author{Caleb Whittier}
\affiliation{Department of Materials Science and Engineering, McMaster University, Hamilton, Ontario L8S 4L8, Canada.}

\author{Joseph Spielbauer}
\affiliation{Department of Physics \& Astronomy, University of Iowa, Iowa City, IA 52242, United States.}

\author{Nabil D. Bassim}
\affiliation{Department of Materials Science and Engineering, McMaster University, Hamilton, Ontario L8S 4L8, Canada.}
\affiliation{Canadian Centre for Electron Microscopy, McMaster University, Hamilton, Ontario L8S 4L8, Canada.}

\author{Thomas G. Folland}

\author{Ravitej Uppu}
\email{ravitej-uppu@uiowa.edu}
\affiliation{Department of Physics \& Astronomy, University of Iowa, Iowa City, IA 52242, United States.}

%\keywords{Defect centers, Quantum emitters, Two-dimensional materials, Hexagonal boron nitride, Focused ion beam, Ion irradiation}

\begin{abstract}
Ion irradiation creates optically active defects in wide-bandgap van der Waals materials, yet most approaches tune defect formation by varying the ion species, energy, or fluence while leaving the incidence geometry fixed.
The ion-incidence angle is established here as a geometric control parameter for engineering visible emitters in hexagonal boron nitride (hBN).
The angle and ion fluence of a plasma-focused heavy-ion (Xe$^+$) beam are varied across hBN flakes of different thickness, and the resulting photoluminescence is quantified. 
In thick flakes, oblique irradiation shifts the fluence for maximum emission by nearly two orders of magnitude relative to normal incidence, whereas thin flakes exhibit an angle-independent optimum.
Ion-trajectory simulations attribute this thickness dependence to lateral redistribution of the collision cascade and enhanced oblique sputtering.
Atomic force microscopy identifies distinct processing regimes that delineate the useful defect-creation window.
Post-irradiation annealing quenches the emission and shifts the spectral weight toward the green-yellow band while preserving the angle-dependent activation trends.
Spectrally resolved lifetime measurements show comparable biexponential dynamics for normal and oblique incidence, consistent with emission from related defect families rather than a geometry-specific emitter species. 
These results establish ion-incidence geometry as a materials-level knob for programming optical defect activation and spatial defect distributions in van der Waals photonic materials.
\end{abstract}

\maketitle

%%%%%%%%%%%%%%%%%%%%%%%%%%%%%%%%%%%%%%%%%%%%%%%%%%%%%%%%%%%%%%%%%%%%%%%%%%%%%%%%%%%%%%%%%%%%
%%%%%%%%%%%%%%%%%%%%%%%%%%%%%%%%%%%%%%%%%%%%%%%%%%%%%%%%%%%%%%%%%%%%%%%%%%%%%%%%%%%%%%%%%%%%

\section{Introduction}

Optically active defects in wide-bandgap materials underpin a growing range of nanophotonic and quantum-information technologies \cite{weber2010quantum, atature2018material, Wolfowicz2021}, in which the localized defect states provide the optical functionality required to generate, manipulate, or read out quantum states \cite{bathen2021manipulating, ccakan2025quantum, Bassett2018}. 
Translating this functionality into devices, however, is as much a materials-processing challenge as a photophysical one: the optical response depends not only on the defect identity, but also on defect density, spatial distribution, and spectral composition \cite{kianinia2022quantum, hou2025engineering}. 
Two-dimensional van der Waals (vdW) materials are attractive in this context because their reduced dimensionality facilitates post-growth defect activation \cite{akkanen2022optical, montblanch2023layered}, while their planar geometry supports photonic integration \cite{elshaari2021deterministic, Li2021, Parto2022, Azzam2023}. 
Among them, hexagonal boron nitride (hBN) is a prominent host for defect-related emission owing to its large band gap, chemical stability, and emitters spanning the visible to the near-infrared \cite{caldwell2019photonics, ccakan2025quantum}.

\begin{figure*}[htpb!]
\centering
\includegraphics[width=1\textwidth]{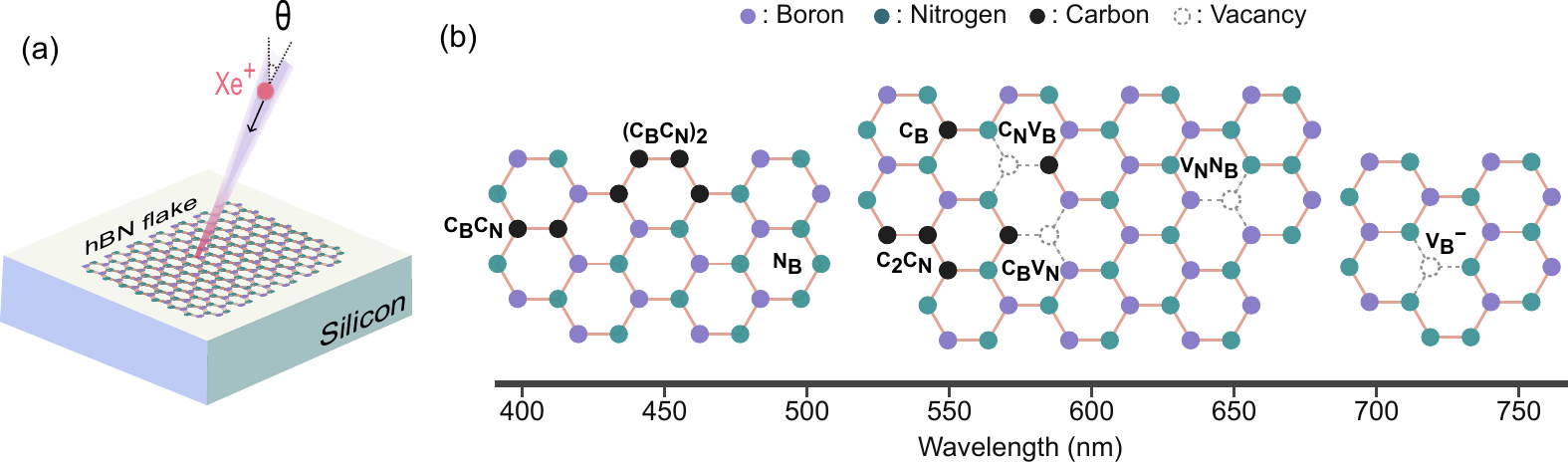}
\caption{\textbf{Ion irradiation scheme and defect-states in hBN.} (a) Schematic of Xe$^+$ ion irradiation on hBN flakes at an incident angle $\theta^{\circ}$. (b) The distribution of vacancy defect states in hBN is illustrated by their atomic lattice configurations relative to their emission wavelengths. Different defect states in hBN with respect to their emission wavelengths.}
\label{fig:fig1}
\end{figure*}

Ion or electron irradiation provides a versatile route to activating and modifying luminescent defects in hBN \cite{toledo2018electron, kianinia2020generation, liang2023high}. 
Focused ion beam irradiation has enabled localized generation of optically active defects in predefined regions \cite{glushkov2022engineering, gale2022site, wu2025site}, where the irradiation fluence quantitatively controls the density of vacancy-related centers \cite{carbone2025quantifying}. 
Typically, irradiation conditions are varied to tune the defect formation or activation \cite{kianinia2020generation, Guo2022, venturi2024selective}, while post-irradiation thermal treatment restructures the resulting defect population \cite{valerius2017annealing, Suzuki2023}. 
The mass of the projectile species and its kinetic energy determine the density and spatial extent of the collision cascade and therefore influence the intensity and spectral composition of the activated emission \cite{hoflich2023roadmap}. 
Because several defect families contribute across the visible spectral range \cite{mendelson2021identifying, huang2022carbon, Cholsuk2024}, irradiation can activate overlapping populations, complicating selective optical engineering (cf., Figure \ref{fig:fig1}) \cite{Qiu2024}. 
Controlling the spatial distribution is likewise important for integration with nanophotonic structures, as the overlap of emitter dipole location and orientation with the spatially varying polarized local density of optical states determines the light-matter interaction efficiency \cite{Lodahl2015, Uppu2020, Ostfeldt2022, sortino2024optically, chowdhury2025resonant}. 
The materials challenge is therefore not simply to create defects, but to control the optical defect landscape generated by irradiation.

Most studies address this challenge by varying ion species, energy, or fluence while retaining a fixed, typically normal-incidence geometry. 
Focused irradiation can define the lateral extent of the exposed region \cite{hennessey2025framework}, but the projected collision cascade remains coupled to the selected projectile, energy, and incidence direction. 
Varying the ion-incidence angle introduces an additional geometric degree of freedom: oblique trajectories modify the lateral and depth-projected distributions of deposited energy and atomic displacements without changing the ion species or energy. 
Incidence angle can also alter sputter yields and erosion rates \cite{wittmaack1990effect}, creating a competition between defect generation and material removal. 
While normal incidence can permit deeper ion propagation along favorable crystallographic directions, oblique incidence suppresses channeling\cite{vantomme201650}, thereby redistributing the collision cascade towards shallower depths and larger lateral areas. 
This near-surface energy deposition in oblique irradiation was leveraged to induce anisotropic damage and surface restructuring in bulk materials \cite{katharria2007self, kamaliya2021tailoring}. 
In graphene, it has been used to control defect generation \cite{bai2015bombarding}, nanopore formation \cite{bai2016nanopore}, and substitutional implantation \cite{bai2016improving}. 
Despite these precedents, incidence geometry has received comparatively little attention as a control parameter for the activation of optical defects in wide-bandgap vdW materials.

Here, we investigate the ion-incidence geometry as an independent control parameter for engineering visible-light-emitting defect activation in hBN. 
By varying the angle of Xe$^+$ in a plasma-focused ion beam (PFIB) irradiation, we reveal that optical activation is governed not only by the number of defects created, but by how the collision cascade is distributed and retained within the material.
The resulting interplay between areal density of defect creation, cascade confinement, and material removal results in a pronounced angle- and thickness-dependent optimal ion fluence for activating visible emitters.
Correlative photoluminescence (PL) spectroscopy and atomic force microscopy (AFM) measurements are used to capture this interplay, while confirming that incidence geometry only preserves the underlying species of emissive defects.
Incidence angle therefore offers a simple geometric handle for tuning irradiation-activated emission without altering the projectile or beam energy, providing a scalable strategy for defect engineering in vdW photonic materials.
%%%%%%%%%%%%%%%%%%%%%%%%%%%%%%%%%%%%%%%%%%%%%%%%%%%%%%%%%%%%%%%%%%%%%%%%%%%%%%%%%%%%%%%%%%%%%%
%%%%%%%%%%%%%%%%%%%%%%%%%%%%%%%%%%%%%%%%%%%%%%%%%%%%%%%%%%%%%%%%%%%%%%%%%%%%%%%%%%%%%%%%%%%%

\section{Results and Discussion}

\subsection{Incidence geometry and flake-thickness-dependent optical defect activation}
\label{sec:activation}

To determine how ion-incidence geometry controls optical defect activation, Xe$^{+}$ PFIB irradiation was performed on mechanically exfoliated hBN flakes on silicon (100) substrates at near-normal ($0\si{\degree}$) and oblique ($30\si{\degree}$ and $60\si{\degree}$) incidence, as illustrated in Figure \ref{fig:fig1}(a). 
Three representative flake thicknesses (40, 130, and 240~nm) were selected for the analysis presented in the main text, with AFM thickness measurements and post-irradiation optical images shown in Figure~S1. 

\begin{figure*}[htpb!]
\centering
\includegraphics[width=1\textwidth]{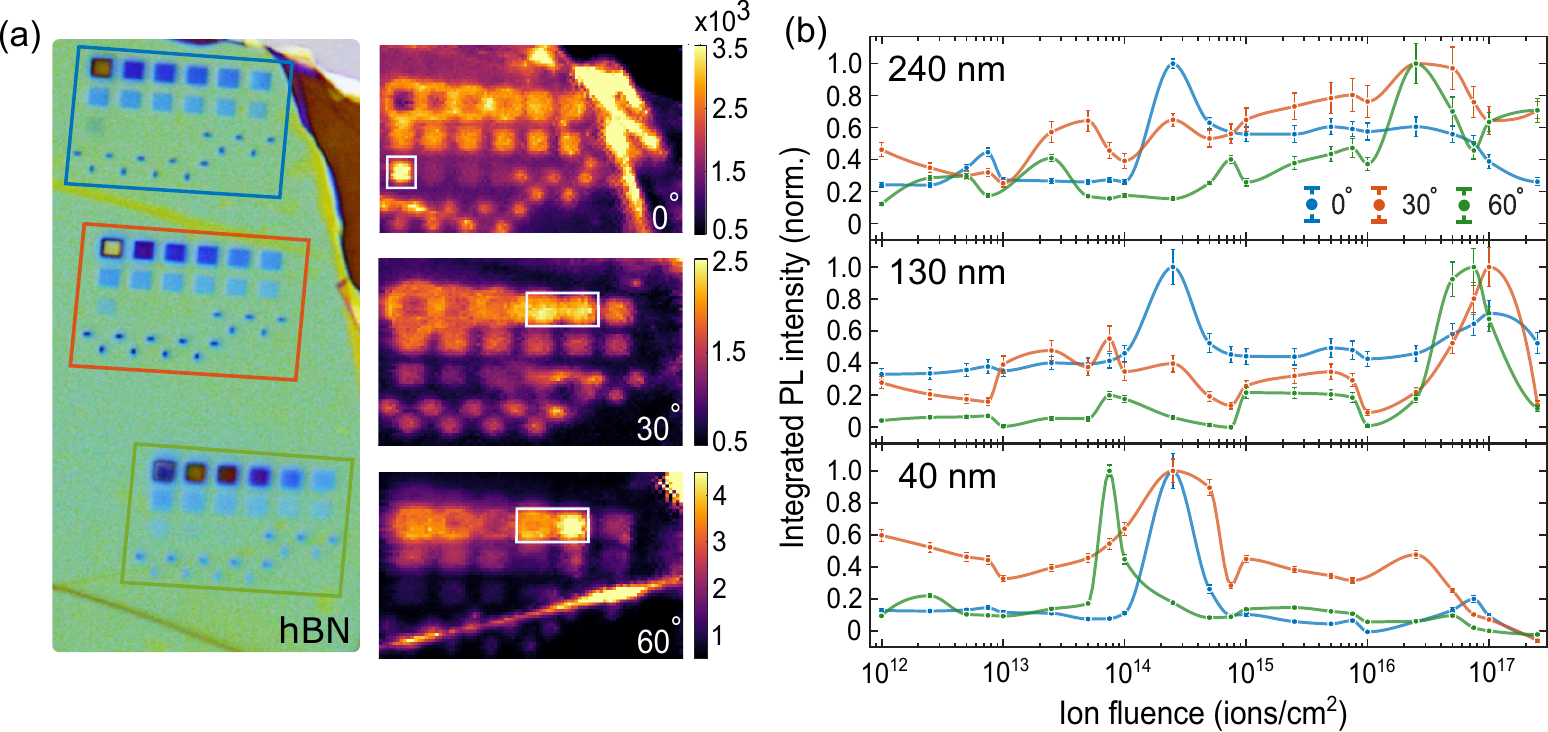}
\caption{\textbf{Angle- and thickness- dependence ion irradiation fluence optimization.} (a) Bright-field optical images (left) and corresponding spatial maps at 566 nm (right) for hBN flakes of thickness 240 nm. Grid outlined in blue, red, and green indicates ion irradiation at 0$^{\circ}$, 30$^{\circ}$, and 60$^{\circ}$ angle incidences, respectively. Each grid contains a series of 2~\si{\micro\meter} × 2~\si{\micro\meter} squares corresponding to different ion fluences arranged sequentially from top to bottom (highest to lowest). The lowest-fluence areas are designated by three reference dots. The brightest ion-activated regions for all three incidence geometries are outlined by white boxes. (b) Normalized average PL intensities as a function of ion fluence are plotted for all flakes at 0$^{\circ}$, 30$^{\circ}$, and 60$^{\circ}$ angle incidence. Data points are calculated by averaging spatial PL intensities across four representative wavelengths and subsequent normalization. Solid lines represent spline interpolations to guide the eye. The error bars represent the standard deviation of the averaged spectral response.}
\label{fig:fig2}
\end{figure*}

On each flake, twenty-two spatially isolated $2 \times 2~\si{\micro\meter}^2$ regions were irradiated for each ion-incidence angle over a fluence range of $10^{12}$--$2.5 \times 10^{17}$~ions/$\si{\centi\meter}^2$. 
The dose in each region was controlled by the exposure time, which was adjusted between 0.6 ms and 160 s (Supplementary Table S1). 
Room-temperature photoluminescence (PL) was measured under continuous-wave 405~nm excitation, with the laser filtered out using a 450~\si{\nano\meter} long-pass spectral filter.
Wavelength-resolved spatial PL maps were acquired using a tunable 2~nm bandpass filter at select wavelengths across the visible band. 
Figure \ref{fig:fig2}(a) shows the bright-field optical microscope image alongside the PL map at 566~nm measured from the 240~nm flake. 
The boxes outlined in blue, red, and green in the microscope image demarcate the irradiation at $0\si{\degree}$, $30\si{\degree}$, and $60\si{\degree}$, respectively. 
Within each grid, the boxes are arranged from the highest fluence at the top left to the lowest at the bottom right, with the three fiducial markers used to identify the low-irradiation-fluence regions. 
Each incidence geometry exhibits a finite activation window: the PL initially increases with fluence, reaches a maximum, and then decreases at higher fluence. 
However, the optimal irradiation fluence that yields the brightest PL (marked by white boxes in the PL image) depends strongly on the incidence angle.
To quantify the dependence of irradiation fluence across different ion incidence geometries, the PL signal integrated over each $2 \times 2~\si{\micro\meter}^2$ irradiated area was used as a measure of optical defect activation, which is robust to local optical inhomogeneity.

Wavelength-resolved spatial PL mapping was carried out on flakes of different thicknesses at four representative wavelengths across the visible spectrum (491~nm, 526~nm, 566~nm, and 623~nm; see SI Figure~S2 for detailed data).
Figure \ref{fig:fig2}(b) shows the ion-fluence-dependent PL for three flakes of different thicknesses (panels) illuminated with ion beams at different incidence geometries.
The data points in the plot correspond to the PL signal averaged across the four wavelengths, with the standard deviation shown as the error bar.
The line connecting the datapoints is a spline interpolation meant solely as a guide to the eye.
For the 240~nm thick flake, normal incidence produces the brightest emission at an ion fluence of $2.5 \times 10^{14}$~ions/$\si{\centi\meter}^2$, whereas the PL maximizes at $30\si{\degree}$ and $60\si{\degree}$ ion incidence occur around $2.5$--$5.0 \times 10^{16}$~ions/$\si{\centi\meter}^2$. 
Oblique irradiation, therefore, requires nearly two orders of magnitude higher fluence at optimal PL brightness. 
Similar irradiation-fluence dependence at normal and oblique incidence is observed in the 130~nm flake.
In contrast, the thin 40~nm flake and an additional 35~nm (Figure~S6) exhibit a largely angle-independent optimum fluence around $10^{14}$~ions/$\si{\centi\meter}^2$ that maximizes PL signal. 
This thickness-dependent bifurcation reveals incidence geometry as a control parameter for the fluence required to activate visible emission. 
The underlying physical mechanisms, including lateral redistribution of the collision cascade and angle-dependent sputtering, are examined through ion-trajectory simulations and AFM measurements in Sections 2.4--2.5. 
%%%%%%%%%%%%%%%%%%%%%%%%%%%%%%%%%%%%%%%%%%%%%%%%%%%%%%%%%%%%%%%%%%%%%%%%%%%%%%%%%%%%%%%%%%%%
%%%%%%%%%%%%%%%%%%%%%%%%%%%%%%%%%%%%%%%%%%%%%%%%%%%%%%%%%%%%%%%%%%%%%%%%%%%%%%%%%%%%%%%%%%%%

\subsection{Spectral composition of the optical defect ensemble}
\label{sec:spectra}

\begin{figure*}[htpb!]
\centering
\includegraphics[width=1\textwidth]{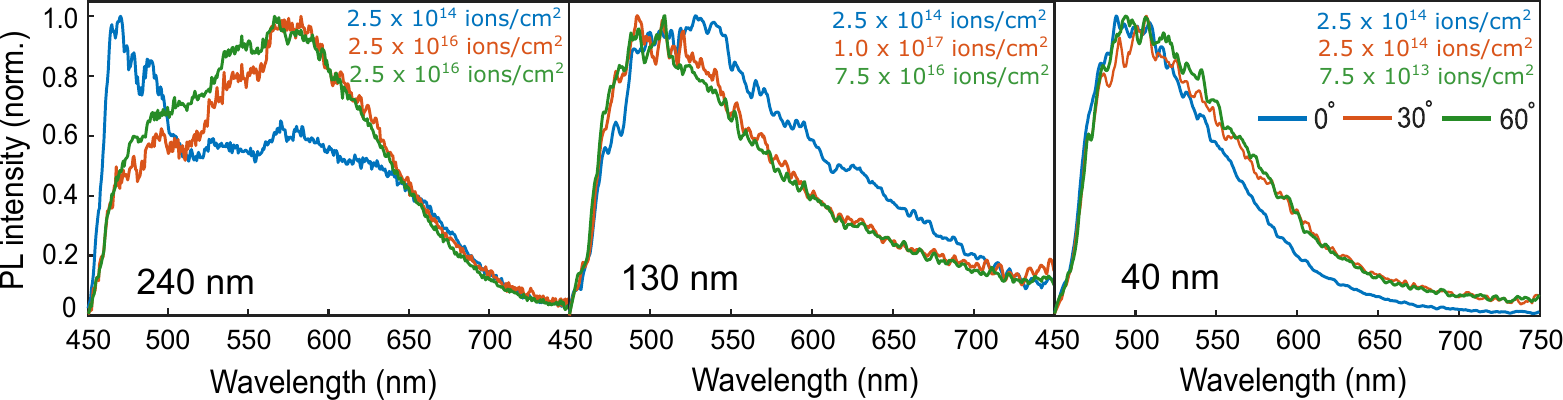}
\caption{\textbf{Photoluminescence spectra at optimized irradiation fluences.} Corresponding optimized ion-fluences for different irradiation conditions are mentioned in the legends.}
\label{fig:fig3}
\end{figure*}

Having established that incidence angle and flake thickness control the fluence required for maximum optical activation (Section~\ref{sec:activation}), we next ask whether the corresponding irradiation conditions produce spectrally distinct emissive populations or redistribute a common one. 
Because the optimal fluence differs by nearly two orders of magnitude between normal and oblique incidence in the thicker flakes, the spectra compared here are necessarily acquired at incidence-geometry-specific optima. 
The effects of incidence angle and absolute fluence are therefore not fully separable, and the lineshape data must be interpreted with this coupling in mind. 
Figure~\ref{fig:fig3} shows the PL spectra measured from irradiation regions that maximize the PL signal, indicated in Figure~\ref{fig:fig2}(b). 
Supplementary Figures~S3--S5 show the comprehensive PL spectra measured at each irradiation fluence and geometry on the three flakes with different thicknesses.

Notably, Figure~\ref{fig:fig3} highlights broadband visible emission spanning $450$--$700$~nm for all irradiation conditions. 
For oblique incidence angle ($30\si{\degree}$ and $60\si{\degree}$), the PL lineshapes are nearly identical for each flake thickness.
On 40~nm and 130~nm thin flakes, the spectral lineshapes under oblique incidence show only a minor redistribution compared to normal incidence geometry, despite the pronounced angle-dependent increase in optimal fluence for the $130$~nm flake. 
The $240$~nm flake shows the largest change in spectral weight, with the oblique-incidence spectra exhibiting higher green-yellow spectral weight at the highest fluences, compared to the dominant blue emission at normal incidence.
This feature is reproducible across ion fluences shown in Supplementary Figure~S3. 
This residual redistribution covaries with the large fluence offset between the geometries and cannot be uniquely assigned to the incidence angle. 
Additionally, a weak oscillatory modulation in this thicker flake is seen at all incidence angles, consistent with thickness-dependent Fabry--P\'erot interference (see Supplementary Figure~S7).

The broad emission spectra measured across all flake thicknesses and irradiation conditions indicate that changes in irradiation geometry do not activate an entirely new type of emitters. 
Instead, the slight shift in the spectral weight with the incidence angle reflects a reorganization of emission across different defect families, summarized in Figure~\ref{fig:fig1}(b). 
Across the visible spectrum, native and carbon-related configurations are anticipated to occur when irradiating commercial hBN crystals under vacuum. 
These are used to identify plausible defect families rather than to assign a specific spectral feature to one microscopic center. 

Direct momentum transfer from Xe$^{+}$ on pristine hBN can create native boron and nitrogen vacancies, antisites, and vacancy--antisite complexes without an extrinsic chemical precursor. 
Native defects proposed to emit within the visible range include $N_B$-related states in the blue and antisite--vacancy complexes such as $N_BV_N$ toward the red \cite{li2025native, Tran2015, Tran2016}. 
The negatively charged boron vacancy $V_B^{-}$ is not expected to contribute substantially since its principal emission lies near $800$~nm \cite{venturi2024selective}. 
The flakes used in our measurements exhibited no discernible visible PL before irradiation, confirming that the observed emission was activated by ion exposure. 
This absence of pre-irradiation PL does not, however, exclude the presence of chemically incorporated impurities that were initially optically inactive.
Alongside the native defects, carbon-related defect complexes can contribute to the observed emission.
Commercial hBN crystals can contain ppm-level carbon impurities introduced during synthesis, while additional contamination may arise during handling and flake transfer.
Such impurities are optically inactive in the as-prepared material but become emissive after irradiation creates nearby vacancies, promotes atomic rearrangement, or modifies the local charge environment. 
Candidate carbon-related emitters span much of the visible spectrum, including the carbon tetramer $(C_BC_N)_2$ in the blue and configurations such as $C_B$, $C_BV_N$, $C_NV_B$, and $C_2C_N$ at longer wavelengths~\cite{maciaszek2024blue, mendelson2021identifying, huang2022carbon, tang2025structured, wu2025site}. 
Our observations are therefore consistent with contributions from both native and carbon-related defect configurations. 
However, because the reported spectral signatures of these centers overlap substantially, the present ensemble measurements do not permit their individual contributions to be resolved.

While the changes in the PL spectra (Figure~\ref{fig:fig3}) indicate that the relative populations of defect configurations shift with incidence angle, their overlapping spectral signatures prevent us from quantifying the restructuring of the populations. 
However, we note that recent atomistic simulations of low-energy carbon implantation into monolayer hBN predict that oblique incidence can enhance the yield of carbon-related color centers \cite{Ren2025}, establishing a precedent for irradiation geometry altering the relative defect populations.
Note that, unlike our Xe$^{+}$ irradiation, this prior work employs carbon as both the projectile and precursor there. 
Resolving the relative population changes requires evidence beyond the PL spectral envelopes, which the subsequent sections supply through time-resolved measurements of emitter identity (Section~\ref{sec:tres}) and post-irradiation annealing of the defect composition (Section~\ref{sec:annealing}).

%%%%%%%%%%%%%%%%%%%%%%%%%%%%%%%%%%%%%%%%%%%%%%%%%%%%%%%%%%%%%%%%%%%%%%%%%%%%%%%%%%%%%%%%%%%%
%%%%%%%%%%%%%%%%%%%%%%%%%%%%%%%%%%%%%%%%%%%%%%%%%%%%%%%%%%%%%%%%%%%%%%%%%%%%%%%%%%%%%%%%%%%%

\subsection{Defect-family invariance across geometries}
\label{sec:tres}

\begin{figure*}[htpb!]
\centering
\includegraphics[width=1\textwidth]{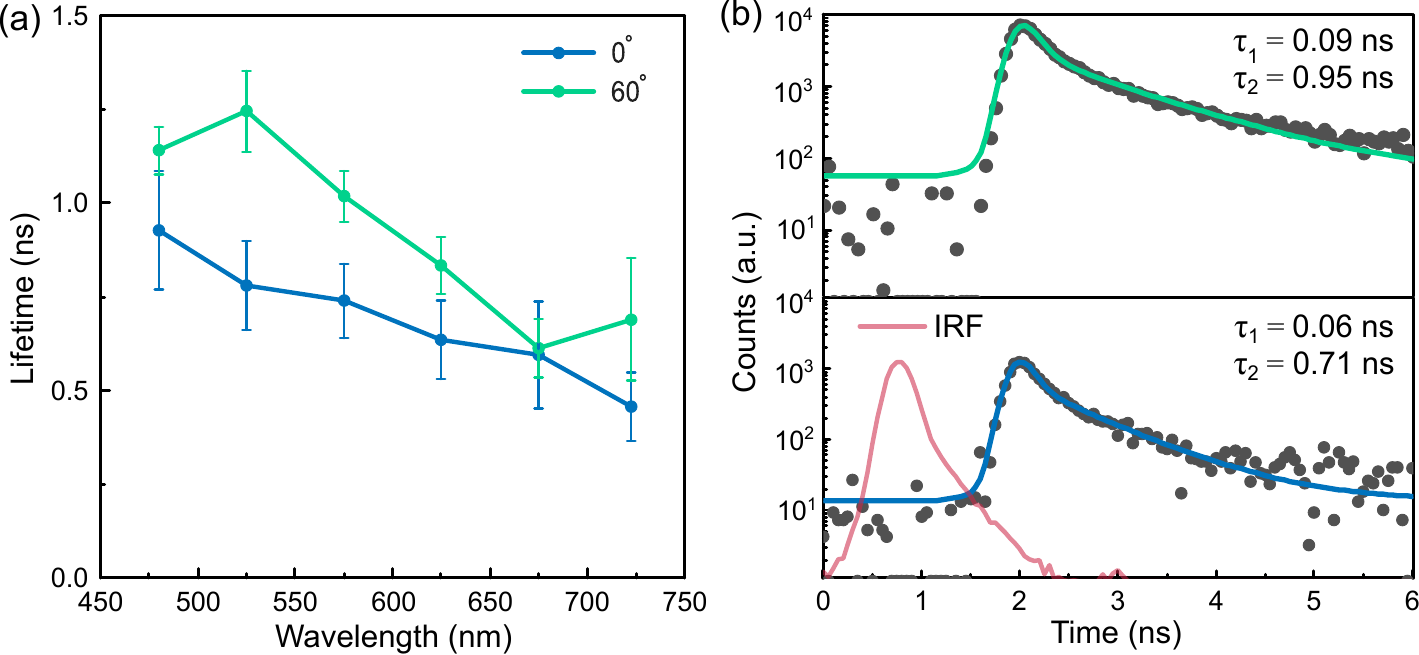}
\caption{\textbf{Spectrally resolved photoluminescence lifetime analysis of hBN under normal and oblique ion incidence.}
(a) Wavelength-dependent PL lifetimes measured for normal incidence (\(0^\circ\), blue) and oblique incidence (\(60^\circ\), green). Error bars indicate the standard deviation of the fitted lifetimes within a \(\pm\SI{25}{\nano\meter}\) spectral window centered at each wavelength, measured using a \(\SI{2}{\nano\meter}\)-bandwidth filter. 
(b) Representative time-resolved PL traces measured at \(\SI{566}{\nano\meter}\) for \(0^\circ\) incidence (bottom) and \(60^\circ\) incidence (top), together with biexponential fits. The extracted fast and slow decay constants, \(\tau_1\) and \(\tau_2\), are indicated in each panel. The red curve shows the measured instrument response function (IRF).}
\label{fig:fig4}
\end{figure*}

Wavelength-dependent time-resolved photoluminescence (TRPL) is measured across the visible spectrum on the optimized-dose regions of the $0\si{\degree}$ and $60\si{\degree}$ sites of the $130$~nm flake. 
Wavelength-dependent spontaneous emission lifetimes are measured by filtering the emission over $460$--$745$~nm using a $2$~nm spectral bandpass filter.
Measured data at each wavelength is fitted to a bi-exponential decay convolved with the instrument response function (IRF), yielding a fast component $\tau_1$ at or near the instrument response limit and a slower component $\tau_2$ that we take as the characteristic emission lifetime. 
IRF was measured by sending attenuated pulses from a spectrally-filtered supercontinuum source tunable across the visible spectrum to capture wavelength-dependent timing jitter of the avalanche photodiode.

Figure~\ref{fig:fig4}(a) shows the mean of the fitted $\tau_2$ values across a $50$~nm span at each central wavelength, corresponding to 25 non-overlapping \SI{2}{\nano\meter} bands; the errorbars represent the standard deviation. 
Across the visible spectrum, both irradiation geometries yield characteristic lifetimes in the same $0.5$--$1.25$~ns range and follow a similar decreasing trend with increasing wavelength (Figure~\ref{fig:fig4}(a)). 
The $60\si{\degree}$ values exceed those at $0\si{\degree}$ values over most of the spectral range, with a maximum difference of about \SI{0.4}{\nano\second} near $525$~nm, before converging at wavelengths approaching $700$~nm. 
Representative decays measured at $566$~nm ($2$~nm bandpass) yield $\tau_2 = 0.95$~ns at $60\si{\degree}$ incidence and $0.71$~ns at $0\si{\degree}$ incidence (Figure~\ref{fig:fig4}(b)), while retaining the same biexponential decay form.
The comparable lifetime range and spectral dependence indicate that both irradiation geometries generate emission from a similar underlying family of defect states rather than from distinct species.
The modest lifetime offset is therefore more naturally attributed to an angle-dependent redistribution of contributions across the defect families.
Taken together, the TRPL results indicate that incidence geometry modifies the density and relative population of a common set of defect states, rather than activating a different species of defect centers.
Next, we investigate the angle- and flake-thickness-dependence of the observed PL signal through ion--matter interaction simulations modeling defect generation.

%%%%%%%%%%%%%%%%%%%%%%%%%%%%%%%%%%%%%%%%%%%%%%%%%%%%%%%%%%%%%%%%%%%%%%%%%%%%%%%%%%%%%%%%%%%%
%%%%%%%%%%%%%%%%%%%%%%%%%%%%%%%%%%%%%%%%%%%%%%%%%%%%%%%%%%%%%%%%%%%%%%%%%%%%%%%%%%%%%%%%%%%%

\subsection{Geometric redistribution of the collision cascade}
\label{sec:cascade}

\begin{figure*}[htpb!]
\centering
\includegraphics[width=1\textwidth]{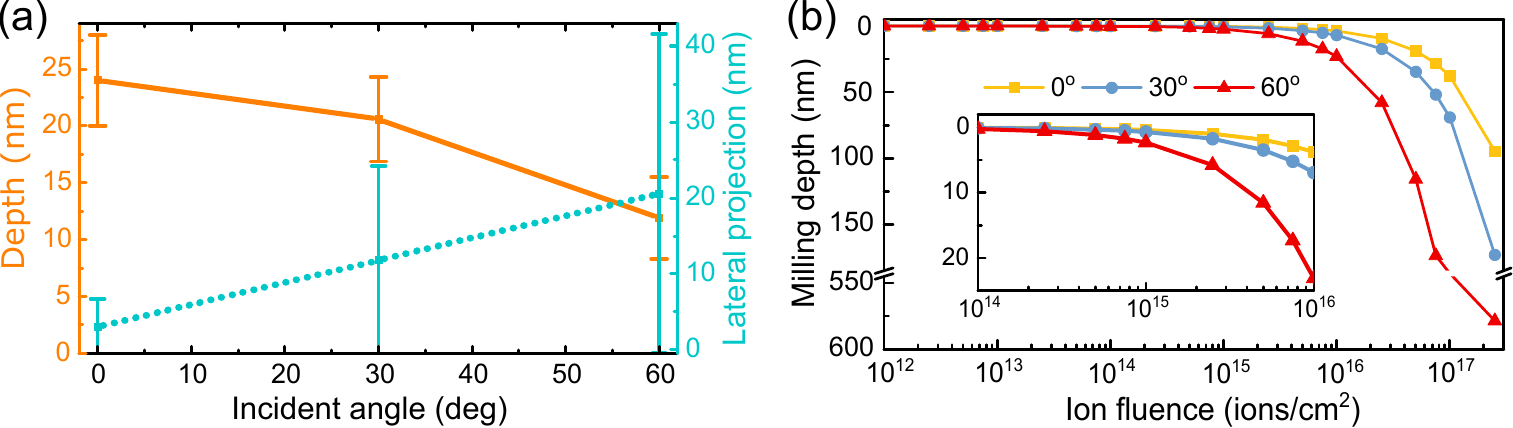}
\caption{\textbf{TRIM analysis of angle-dependent ion irradiation and milling in hBN.} (a) Calculated ion penetration depth (left, orange) and lateral spread (right, cyan) with error bars indicate the corresponding straggle in the hBN as a function of incidence angles. (b) Calculated milling depth as a function of ion fluence for irradiation angles; inset highlights the onset of significant material milling.}
\label{fig:fig5}
\end{figure*}

We modeled Xe$^{+}$ ion--solid interactions using TRIM/Iradina simulations of a layered hBN-on-Si geometry to capture the angle- and thickness-dependent shift of the optimal activation fluence. 
The simulations provide the longitudinal range, the lateral projection and associated straggle, and the sputter yield as functions of ion incidence angle (Figure~\ref{fig:fig5}, Table~S2).
The absolute number of vacancies depends on the threshold displacement energies $E_d$ assigned to B and N atoms. 
We use $E_d(\mathrm{B})=19.36$~eV and $E_d(\mathrm{N})=23.06$~eV, as obtained from first-principles molecular-dynamics calculations for pristine, charge-neutral monolayer hBN \cite{Kotakoski2010, Bui2023}.
Although variations in these values would alter the predicted absolute vacancy densities, the relative angle-dependent lateral and longitudinal dimensions of the collision cascade are expected to be less sensitive to the precise choice of $E_d$.

The calculated vacancy yield varies only weakly with incidence angle, with values of $505.2$, $506.3$, and $502.9$~vacancies/ion at $0\si{\degree}$, $30\si{\degree}$, and $60\si{\degree}$, respectively (Table~S2). 
Thus, within the accuracy of the simulations, tilting the incident beam does not appreciably change the total number of vacancies generated per ion.
Instead, it primarily changes their spatial distribution within the hBN layer. 
As the beam is tilted away from the surface normal, the collision cascade becomes shallower and more laterally extended. 
The longitudinal projected range decreases from $23.8$~nm at $0\si{\degree}$ to $20.7$~nm at $30\si{\degree}$ and $11.9$~nm at $60\si{\degree}$, while the lateral projected range grows from $3.0$~nm to $11.9$ and $20.6$~nm, respectively (Figure~\ref{fig:fig4}(a); Table~S2). 
Normal incidence therefore produces a comparatively deep, longitudinally extended cascade, whereas irradiation at $60\si{\degree}$ produces a shallow cascade distributed over a substantially larger lateral area. 
The lateral and radial straggle are also comparable at oblique incidence, reaching $12.5$~nm at $30\si{\degree}$ and $20.9$~nm at $60\si{\degree}$).
This similarity indicates that the in-plane broadening is approximately isotropic rather than being concentrated predominantly to the beam-tilt direction.
The relevant geometric effect is therefore a two-dimensional dilution of the collision cascade over the flake surface.

In the context of the PL measurements, the confocal PL geometry collects emission throughout the full flake thickness.
For the 0.9-NA objective, the depth of focus of the \SI{405}{\nano\meter} excitation beam exceeds \SI{300}{\nano\meter}, which is greater than the maximum flake thickness of 240~nm investigated here.
The experimentally relevant quantity is therefore the vacancy density integrated through the flake thickness and projected onto the surface plane.
Approximating the projected area of the cascade as $A_{XY}\sim \pi (\Delta R_{\text{lat}})^2/\cos\theta$, where \(\Delta R_{\mathrm{lat}}\) is the lateral extent and \(\theta\) is the incidence angle, gives projected areal vacancy densities of $1.2\times10^{15}$~cm$^{-2}$, $8.9\times10^{13}$~cm$^{-2}$, and $1.8\times10^{13}$~cm$^{-2}$ at $0\si{\degree}$, $30\si{\degree}$, and $60\si{\degree}$, respectively (Supplementary Table~S2). 
Relative to normal incidence, this corresponds to reductions in the areal defect density generated per ion by factors of approximately \(13\) at \(30\si{\degree}\) and \(64\) at \(60\si{\degree}\).
Note that this reduction does not displace the ions outside the irradiated boxes or the optically probed area. 
Even at \(60\si{\degree}\), the calculated lateral extent of the cascade, \(\Delta R_{\mathrm{lat}}\approx21\)~nm, remains much smaller than both the \SI{2}{\micro\meter} irradiation box and the approximately \SI{500}{\nano\meter} optical collection diameter. 
Instead, the same number of vacancies is distributed over a larger local area, lowering the through-thickness-integrated defect density produced by each ion. 
A correspondingly larger fluence is therefore required to reach a comparable local density of optically active centers.

The geometric dilution provides a natural explanation for the observed shift in the ion fluence required to maximize PL signal at oblique incidence (Figure~\ref{fig:fig2}(b)). 
In particular, the calculated reduction in areal vacancy density at \(60\si{\degree}\) is of the same order as the measured shift in the activation fluence.
At \(30\si{\degree}\), however, the calculated reduction of approximately \(13\) is smaller than the nearly two-order-of-magnitude optimal fluence shift inferred from the PL measurements. 
Geometric redistribution therefore accounts for the principal trend but does not, by itself, determine the complete activation window. 

\begin{figure*}[htpb!]
\centering
\includegraphics[width=0.8\textwidth]{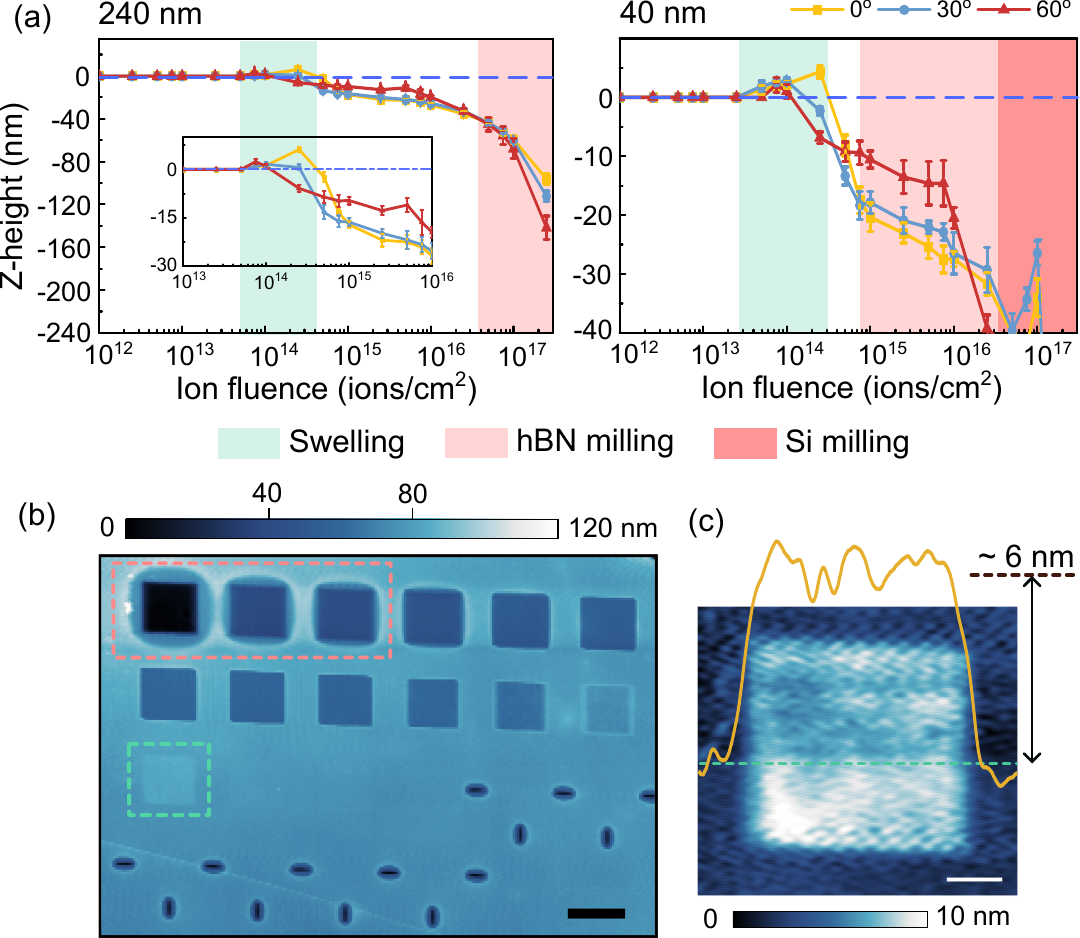}
\caption{\textbf{Thickness- and angle-dependent morphology of ion-irradiated hBN.}
(a) Average AFM-measured height ($Z$) as a function of Xe-ion fluence for hBN flakes with thicknesses of \(240\)~nm (left) and \(40\)~nm (right), irradiated at incidence angles of \(0^{\circ}\), \(30^{\circ}\), and \(60^{\circ}\). The shaded regions indicate swelling (green), sputter-driven hBN thinning (light red), and milling of the underlying Si substrate (dark red). The inset magnifies the swelling-dominated fluence range for the \(240\)~nm flake. Each point represents the mean \(Z\)-height of an individual irradiated box, and the error bars denote the corresponding standard deviation.
(b) AFM topography of the \(240\)~nm hBN flake irradiated at \(0^{\circ}\). Heavily milled regions are outlined by the red dashed box, while the swollen region is marked by the green dashed box. The scale bar corresponds to \(2~\mu\mathrm{m}\).
(c) Magnified AFM image of the swollen region highlighted in (b), together with the corresponding height profile extracted along the green dashed line. The profile reveals a height increase of approximately \(6\)~nm relative to the surrounding unirradiated hBN surface. The scale bar corresponds to \(500\)~nm.
}
\label{fig:fig6}
\end{figure*}

A second contribution arises from angle-dependent material removal.
The simulated sputter yield increases from $3.68$~atoms/ion at $0\si{\degree}$ to $6.91$~atoms/ion at $30\si{\degree}$ and $22.8$~atoms/ion at $60\si{\degree}$ (Table~S2), resulting in a more rapid increase in calculated milling depth at larger incidence angles (Figure~\ref{fig:fig5}(b)).
The milling depth $D$ was estimated from the simulated sputter yield $Y$ according to the relation $D = Y\Phi/\rho$, with $\Phi$ the fluence and $\rho = 1.02\times10^{23}$~atoms/cm$^3$ the atomic density of hBN ($2.1$~g/cm$^3$). 
The calculated milling depth rises sharply with incidence angle (Figure~\ref{fig:fig4}(b)). 
At $\Phi = 2.5\times10^{17}$~ions/cm$^2$, the milling depth is approximately 90~nm at $0\si{\degree}$, $170$~nm at $30\si{\degree}$, and $570$~nm at $60\si{\degree}$, with a measurable onset near $\sim 10^{15}$~ions/cm$^2$.
Thus, higher fluences required for compensating the geometric dilution at oblique ion incidence also produce progressively stronger erosion of the near-surface region. 
Geometric dilution and sputtering play distinct but complementary roles. 
Geometric dilution reduces the areal defect density generated per incident ion and therefore shifts the activation threshold toward higher fluence. 
Sputter-driven removal limits the fluence range over which the damaged layer and its associated emitters can accumulate.
The observed PL behavior in Figure~\ref{fig:fig2}(b) suggests that at $30\si{\degree}$ incidence, the additional influence of material removal contributes to the larger-than-predicted shift in the PL optimum. 
At \(60\si{\degree}\), geometric dilution provides the dominant shift, while the strongly enhanced sputter yield constrains the upper end of the activation window. 

The calculated penetration and milling depths also provide insight into the observed flake-thickness dependence (Figure~\ref{fig:fig2}(b)). 
In the thinnest flakes, the hBN thickness is comparable to the ion-penetration depth, and at elevated fluence, to the calculated milling depth (Figure~\ref{fig:fig5}). 
Incident ions can therefore reach the hBN--Si interface, while progressive milling reduces the amount of hBN available to retain the collision cascade. 
Backscattered ions and substrate-generated recoils may consequently contribute to defect formation within the hBN and reduce the sensitivity of the integrated response to the initial incidence angle. 
This interpretation is consistent with the nearly angle-independent optimal fluence of approximately \(10^{14}\)~ions~cm\(^{-2}\) observed for the thinnest flakes.
In thicker flakes, by contrast, the incident ions are predominantly stopped within the hBN, and the collision cascade is contained within the flake. 
The angle-dependent lateral redistribution is therefore preserved, producing the pronounced angle dependence of the optimal fluence. 
At still higher fluence, material removal becomes substantial and ultimately limits the retained defect population.

%%%%%%%%%%%%%%%%%%%%%%%%%%%%%%%%%%%%%%%%%%%%%%%%%%%%%%%%%%%%%%%%%%%%%%%%%%%%%%%%%%%%%%%%%%%%
%%%%%%%%%%%%%%%%%%%%%%%%%%%%%%%%%%%%%%%%%%%%%%%%%%%%%%%%%%%%%%%%%%%%%%%%%%%%%%%%%%%%%%%%%%%%

\subsection{Observing the morphological process regimes}
\label{sec:morphology}

We used atomic force microscopy (AFM) to identify the morphological regimes associated with defect activation, hBN thinning, and eventual substrate milling, as identified through TRIM simulations in the previous section.
Figure~\ref{fig:fig6}(a) shows the measured $Z$-height  as a function of the ion fluence across irradiation geometries for the 240~nm and 40~nm thick flakes. 
The plotted $Z$-height is the mean relief across the \SI{2}{\micro\meter}$\times$\SI{2}{\micro\meter} irradiated region relative to an unirradiated region, with the error bars corresponding to the standard deviation.
The measured $Z$-height closely follows the same trend as the simulated milling depth $D$ in Figure~\ref{fig:fig5}(b) (Supplementary Figure~S9 for direct comparison), with some evident differences.
Firstly, while simulations predict the onset of appreciable milling near \(10^{15}\)~ions/$\si{\centi\meter}^2$, the AFM measurements reveal morphological changes beginning near \(10^{14}\)~ions/$\si{\centi\meter}^2$.
This difference is expected because the simulations treat hBN as a bulk amorphous target and do not capture its layered structure, defect-assisted erosion, or irradiation-induced reconstruction.

Another interesting feature evident around a fluence of \(10^{14}\)~ions/$\si{\centi\meter}^2$ is the swelling of hBN flakes, irrespective of their thickness.
Thus, the AFM measurements identify three morphological regimes for the \SI{240}{\nano\meter}-thick hBN flake. 
Figure~\ref{fig:fig6}(b) shows a representative region on the 240~nm flake irradiated at normal incidence (see Supplementary Figure~S8 for scans of other regions and flakes).
At low fluence, the irradiated surface remains nearly unchanged. 
Near \(10^{14}\)~ions/$\si{\centi\meter}^2$, the surface exhibits a small positive height change associated with irradiation-induced swelling. 
At higher fluence, the \(Z\)-height decreases progressively as sputter-driven thinning becomes dominant. 
At \(2.5\times10^{17}\)~ions/$\si{\centi\meter}^2$, the measured removal depths are approximately \(90\), \(112\), and \(140\)~nm for incidence angles of \(0\si{\degree}\), \(30\si{\degree}\), and \(60\si{\degree}\), respectively.
Note that these milling depths are smaller than the predicted values, with the disparity increasing with incidence angle.
The \(40\)~nm-thick flake exhibits a markedly narrower processing window because its thickness is comparable to the ion penetration and milling length scales. 
After the swelling around a fluence of \(10^{14}\)~ions/$\si{\centi\meter}^2$, pronounced thinning begins near \(10^{15}\)~ions/$\si{\centi\meter}^2$, and at the highest fluences the hBN is fully removed, followed by milling of the underlying Si substrate. 
This complete removal accounts for the loss of PL at the highest fluences (Supplementary Figure~S4).

Comparing the spatial PL map (Figure~\ref{fig:fig2}(a)) and the AFM image reveals that the PL maximum corresponds to the ion fluence resulting in hBN surface swelling.
A magnified AFM scan (Figure~\ref{fig:fig6}(c)) shows that the hBN surface swells by around 6~nm relative to the unirradiated surface.
We attribute this swelling to irradiation-induced lattice disorder, strain, and the migration of displaced atoms toward the surface \cite{Ren2023, grosso2017tunable}. 
This swelling regime corresponds to the ion fluence that creates a substantial density of defect centers before amorphization and sputtering-driven material removal becomes dominant.
Similar swelling is also observed under oblique irradiation, but with a smaller relief.
The larger sputter yield at oblique angles likely suppresses the accumulation of a swollen near-surface layer by removing damaged material more rapidly.
Taken together, the AFM measurements and the TRIM calculations identify a competition between defect formation and material removal: geometric dilution shifts the activation threshold to higher fluence, while angle-dependent sputtering limits the upper end of the useful processing window.

%%%%%%%%%%%%%%%%%%%%%%%%%%%%%%%%%%%%%%%%%%%%%%%%%%%%%%%%%%%%%%%%%%%%%%%%%%%%%%%%%%%%%%%%%%%%
%%%%%%%%%%%%%%%%%%%%%%%%%%%%%%%%%%%%%%%%%%%%%%%%%%%%%%%%%%%%%%%%%%%%%%%%%%%%%%%%%%%%%%%%%%%%

\subsection{Post-irradiation annealing reveals geometry-dependent defect selection}
\label{sec:annealing}

\begin{figure*}[htpb!]
\centering
\includegraphics[width=0.7\textwidth]{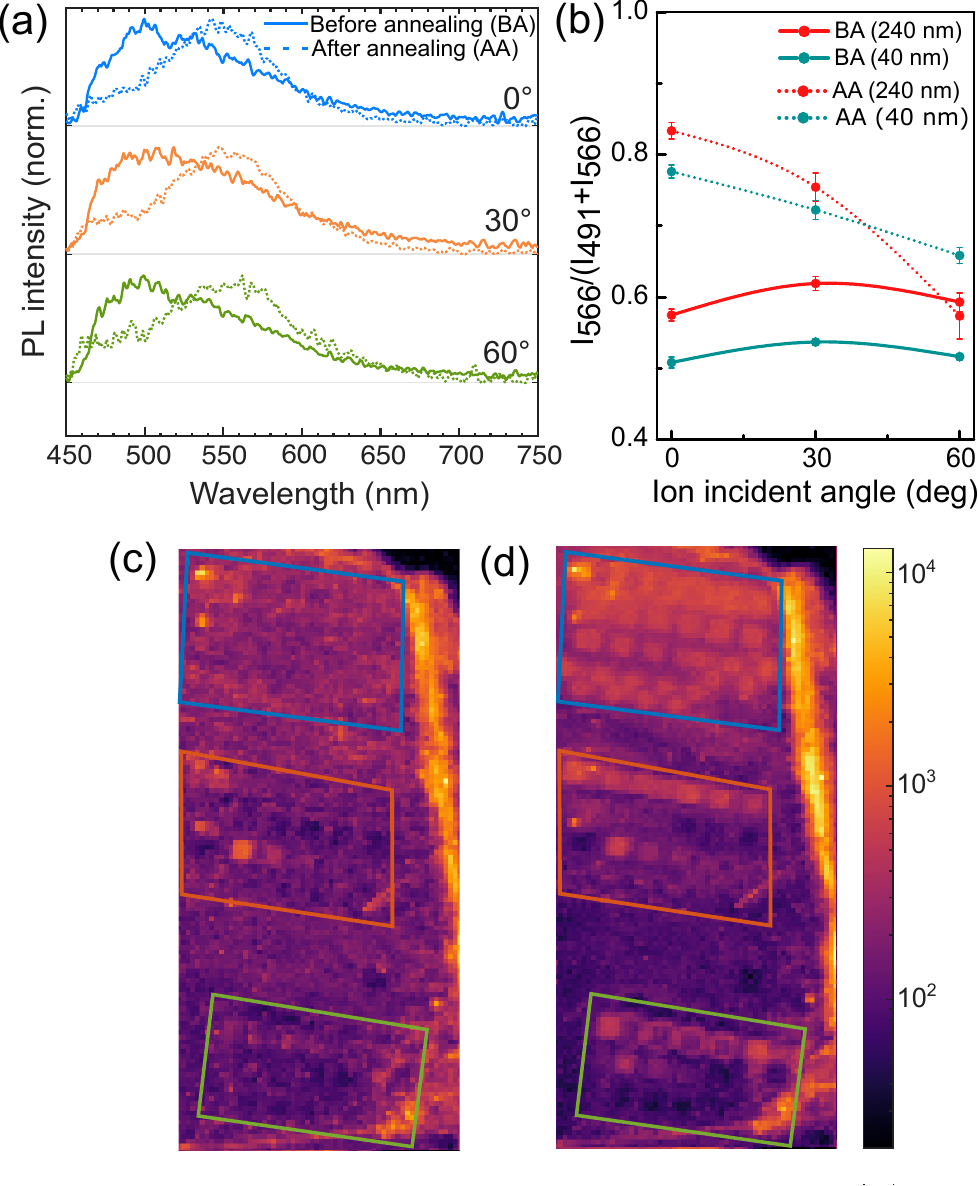}
\caption{\textbf{Effect of post-irradiation annealing.} (a) Normalized PL spectra from the 40 nm thin sample for 0$^{\circ}$, 30$^{\circ}$, and 60$^{\circ}$ angle incidences from top to bottom, respectively, before (dashed) and after (solid) annealing, highlighting a shift in emission weight from cyan to green-yellow channel. (b) The relative spectral weight between 491 nm and 566 nm emission for these channels is calculated,  with respect to the incidence angles for both 240 nm thick (red) and 40 nm thin (green) samples. The solid and dotted lines represent the before-annealing (BA) and after-annealing (AA) conditions, respectively. (c,d) Spatially integrated PL maps centered at 491 nm and 566 nm with a 45 nm bandwidth of the 240 nm thick flake collected after thermal annealing. The blue, red and green boxes represent the 0$^{\circ}$, 30$^{\circ}$, and 60$^{\circ}$ ion irrdiated areas respectively.} 
\label{fig:fig7}
\end{figure*}

Section~\ref{sec:tres} established that the emitters form a common defect family across geometries. 
Post-irradiation annealing shows that geometry nonetheless biases the relative populations within that family. 
The irradiated flakes were annealed at $800\si{\degree}$C for $2$~h under N$_2$, which quenches the integrated visible emission by approximately an order of magnitude under matched PL acquisition conditions. 
This substantial loss in emission is consistent with the annealing-driven recombination, passivation, or reconstruction of a large fraction of the irradiation-induced defect population \cite{Ren2023, venturi2024selective}. 

Figure~\ref{fig:fig7}(a) shows PL spectra from the 240~nm flake at fixed fluence before and after annealing, individually normalized to their peaks to expose the spectral redistribution. 
Annealing selectively preserves the green--yellow emission centered at 560~nm. 
Quantifying this spectral distribution through the spectral fraction $I_{566}/(I_{491}+I_{566})$ (Figure~\ref{fig:fig7}(b)), the pre-annealing value sits near $0.5$--$0.6$ with only a modest angle dependence for both the 40 and 240~nm flakes.
After annealing, the spectral fraction increases to $0.8$ at normal incidence and decreases monotonically toward $0.6$ at $60\si{\degree}$ ion incidence. 
The cyan contribution is therefore preferentially depleted relative to the green--yellow emission, and the degree of this differential depletion depends systematically on the incidence angle.

Because the emitters share a common family, this angle-dependent spectral redistribution indicates that the three irradiation geometries produce different relative populations within it, exposed by
annealing even where the as-irradiated spectra are similar. 
We interpret annealing as a kinetic filter: less stable configurations are annihilated, reconstructed, or rendered nonradiative, while the green--yellow-emitting configurations survive at a higher fraction. 
The redistribution, therefore, reflects preferential preservation rather than absolute enhancement of the 566~nm band. 
Carbon-containing vacancy complexes are a plausible contributor to the PL signal on the annealed samples \cite{mendelson2021identifying, Kumar2023}.
In this picture, the angle dependence reflects geometry-dependent differences in the initial balance of vacancy--impurity configurations, which were obscured before annealing but exposed by their differing thermal stabilities.

The spatially resolved response consolidates this into a geometry-dependent shift of the activation window. 
Figure~\ref{fig:fig7}(c,d) show the PL maps at 491~nm and 566~nm respectively of the 240~nm flake ($0\si{\degree}$, $30\si{\degree}$, $60\si{\degree}$ as blue, red, green boxes).
Cyan (491~nm) emission is broadly suppressed across all fluences and geometries. 
At normal incidence, the broad defect activation window spanning $\sim\!10^{14}$ to $\sim\!10^{16}$~ions/cm$^2$ observed before annealing is largely retained at reduced intensity. 
Comparing with the as-irradiated map (Figure~\ref{fig:fig2}(a)), the optimum fluence at $30\si{\degree}$ ion incidence shifts from $\sim\!10^{16}$~ions/cm$^2$ before annealing to $\sim\!10^{14}$~ions/cm$^2$ afterward,
whereas the optimum under $60\si{\degree}$ ion incidence remains near $\sim\!10^{16}$~ions/cm$^2$.
This difference across oblique incidence angles follows from the observed angle-dependent sputter yield in ion-matter interactions. 
As discussed in Sections~\ref{sec:cascade} and \ref{sec:morphology}, irradiation-induced milling occurs at lower fluence as the irradiation angle increases.
Taken together, the annealing measurements show that ion-incidence geometry biases the fluence-dependent formation and destruction of the activated ensemble: angle sets the dose for maximum initial emission, the cyan/green--yellow weighting after annealing, and the window over which thermally persistent emission remains.

%%%%%%%%%%%%%%%%%%%%%%%%%%%%%%%%%%%%%%%%%%%%%%%%%%%%%%%%%%%%%%%%%%%%%%%%%%%%%%%%%%%%%%%%%%%%
%%%%%%%%%%%%%%%%%%%%%%%%%%%%%%%%%%%%%%%%%%%%%%%%%%%%%%%%%%%%%%%%%%%%%%%%%%%%%%%%%%%%%%%%%%%%

\section{Conclusions}

We have demonstrated ion-incidence geometry as a materials-level control parameter for activating visible defect emission in hBN.
By varying the Xe$^+$ irradiation angle, ion fluence, and flake thickness, we show that the optimal activation window depends strongly on the projected damage geometry.
In thick hBN flakes, oblique irradiation shifts the fluence required for maximum visible PL by nearly two orders of magnitude relative to normal incidence, whereas thin flakes exhibit a much weaker angle dependence.
TRIM simulations and AFM morphology measurements show that this behavior arises from the interplay between the depth-projected ion range, lateral damage spreading, defect accumulation, and sputter-driven removal of the hBN host.
Spectrally resolved PL and lifetime measurements are consistent with incidence angle controlling the effective density and relative populations of related optically active defect configurations, rather than producing an entirely distinct emissive species.
Post-irradiation annealing further quenches the total PL while selectively preserving the green--yellow contribution, revealing geometry-dependent differences within the emitter population that are largely obscured in the as-irradiated envelope spectra.

Together, these results establish oblique ion irradiation as a simple, mask-free route for programming optical defect landscapes in van der Waals photonic materials.
More broadly, incidence-angle control provides a geometric processing knob for tuning the density of activated emitters, as well as the depth and lateral extent of the damaged volume and the spectral balance of the resulting emission.
These parameters are important for controlling the interaction of color centers with planar photonic architectures.

Two directions follow directly from this work.
First, while our measurements establish that incidence geometry alters the relative populations within a common defect family, the microscopic identities of the geometry-selected configurations remain to be resolved.
Correlative single-defect spectroscopy and first-principles assignment of candidate native and carbon-related centers \cite{mendelson2021identifying, maciaszek2024blue, Cholsuk2024} would connect the observed spectral redistribution to specific atomic structures.
Second, combining incidence-angle control with site-selective irradiation could enable individual quantum emitters to be positioned laterally and confined within selected depth ranges \cite{glushkov2022engineering, gale2022site, wu2025site, Stewart2021}.
Such control would provide a route to optimizing emitter overlap with the projected local density of optical states in planar photonic architectures \cite{Lodahl2015,sortino2024optically}.
Together, these steps would extend geometric activation from ensemble-level control toward the deterministic, site- and depth-resolved defect placement required for integrated quantum photonics.

%%%%%%%%%%%%%%%%%%%%%%%%%%%%%%%%%%%%%%%%%%%%%%%%%%%%%%%%%%%%%%%%%%%%%%%%%%%%%%%%%%%%%%%%%%%%
%%%%%%%%%%%%%%%%%%%%%%%%%%%%%%%%%%%%%%%%%%%%%%%%%%%%%%%%%%%%%%%%%%%%%%%%%%%%%%%%%%%%%%%%%%%%

\section*{Methods}

\subsection*{Sample Preparation and Annealing} The hBN flakes were mechanically exfoliated from bulk hBN crystals and transferred to a silicon substrate, with careful attention paid to minimize adhesive residue during the transfer process. 
From the exfoliated flakes, we selected relatively large flakes (lateral size $>$30~\si{\micro\meter}) with thickness ranging from 35~\si{\nano\meter} to 240~\si{\nano\meter}, determined by atomic force microscopy (AFM). 
These flakes were irradiated with Xe$^+$ ion beam using a Thermo Fisher Scientific Helios 5 plasma focused ion beam (PFIB)/ scanning electron microscopy (SEM) dual beam system. 
The accelerated voltage and beam current were set to 30~\si{\kilo\volt} and 10~\si{\pico\ampere}. 
For dose optimization across different incident ion angles, we patterned several $2\times2~\si{\micro\meter}^2$ regions in a two-dimensional array with a pitch of 3~\si{\micro\meter}.
The ion irradiation dose was varied in logarithmic steps over five orders of magnitude across the 22 exposure regions at each incidence angle by adjusting the exposure time.
After PFIB processing, the ion-irradiated regions were imaged by scanning electron microscopy (SEM) at 5 kV to avoid electron-induced damage and further characterized by AFM in non-contact mode.  
For thermal annealing, the samples were treated in a tube furnace (MTI RTP-1000D4) at 800~\si{\celsius} for 2~\si{hours} in 1 atm of N$_2$, with heating and cooling rates of \(\sim \) 40~\si{\celsius}/min. 

\subsection*{Optical Characterization} PL measurements were performed using a home-built confocal microscope. 
Collimated laser light from a 405~\si{\nano\meter} continuous-wave laser (Vortran Stradus 405) was expanded to a diameter of 5~\si{\milli\meter} and imaged onto the back focal plane of a high numerical-aperture objective (Nikon E Plan 100x, NA=0.9) using a 4$f$ relay composed of two lenses ($f=150~\si{\milli\meter}$).
The objective focuses the excitation beam to a diffraction-limited spot ($\approx300~\si{\nano\meter}$) on the sample, which is mounted on piezo scan stages (Newport CONEX-SAG-LS16P).
Fluorescence emission was collected with the same objective and separated from the excitation path using a 450~\si{\nano\meter} dichroic beam splitter (Edmund Optics 69-898).
Residual scattered laser light was further suppressed using a 450~\si{\nano\meter} long-pass filter (Thorlabs FELH450).
The filtered fluorescence was coupled into a 5~\si{\micro\meter} core optical fiber (Thorlabs 780HP) using a 19~\si{\milli\meter} focal length achromatic fiber collimator.
The fiber acts as a spatial pinhole, corresponding to an effective collection diameter of approximately 500~\si{\nano\meter} on the sample.
The collected emission was spectrally filtered using a tunable bandpass filter (2~\si{\nano\meter} bandwidth) and detected with a single-photon avalanche diode (Hamamatsu C16533-050GD) to construct photoluminescence spectra.

For time-resolved PL measurements, the continuous-wave excitation laser was replaced by a pulsed laser source delivering sub-picosecond pulses at 405 nm with an 80 MHz repetition rate, generated via second-harmonic conversion of a Ti:sapphire laser (Spectra-Physics Tsunami). 
Photon arrival times and the corresponding laser synchronization signal were recorded using high-resolution time-tagging electronics (qutools QuTAG HR).
Time-correlated single-photon counting (TCSPC) analysis was performed by fitting the measured time-resolved PL curves with single- or biexponential decay models convoluted with the instrument response function (IRF). 
The IRF was independently measured across the visible spectral range using a tunable, spectrally filtered supercontinuum source generated from the same excitation laser. 
For this measurement, the excitation beam was reflected from a mirror in place of the sample and detected via the same optical and detection pathways. 
The experimentally measured IRF was used in all fits to account for the finite temporal resolution of the system.

%%%%%%%%%%%%%% Acknowledgements %%%%%%%%%%%%%%
\begin{acknowledgments}
This material is based upon work supported by the Office of the Under Secretary of Defense for Research and Engineering under award number FA9550-25-1-0273.
N.D.B. thanks the Natural Sciences and Engineering Research Council of Canada (NSERC) for financially supporting this work under the Alliance International Catalyst Quantum grants program (ALLRP 580935 - 22). 
The authors acknowledge support for carrying out the focused ion beam (FIB) irradiation at the Canadian Centre for Electron Microscopy (CCEM), a national facility supported by McMaster University, the Ontario Research Fund (ORF), and the Canada Foundation for Innovation (CFI).
\end{acknowledgments}
%%%%%%%%%%%%%%%%%%%%%%%%%%%%%%%%%%%%%%%%%%%%%%%%%%%%%%

\bibliography{refs_cleaned}
%%%%%%%%%%%%%%%%%%%%%%%%%%%%%%%%%%%%%%%%%%%%%%%%%%%%%%

\end{document}